# Empowering Enterprise Development by Building and deploying Admin Dashboard using Refine Framework

Gajjala Sai Teja [1], Devi Deepak Manchala [2], Gummadelly Bhargav [3], Mrs. K. Naga Sailaja[4]

Department of ECM, Sreenidhi Institute of Science and Technology, Hyderabad, India

[1]saitejagajjala027@gmail.com;  [2]devideepak109@gmail.com; [3]bhargavgummadelly@gmail.com, [4]nagasailajak@sreenidhi.edu.in

*Abstract*— **This project proposes the development of an advanced admin dashboard tailored for enterprise development, leveraging the Refine framework, Ant Design, and GraphQL API. It promises heightened operational efficiency by optimizing backend integration and employing GraphQL's dynamic data subscription for real-time insights. With an emphasis on modern aesthetics and user-centric design, it ensures seamless data visualization and management. Key functionalities encompass user administration, data visualization, CRUD operations, real-time notifications, and seamless integration with existing systems. The deliverable includes a deployable dashboard alongside comprehensive documentation, aiming to empower enterprise teams with a cutting-edge, data-driven solution.**

**Keywords: Refine framework, Ant Design, Graph QL API, Admin Dashboard, Enterprise Development, Data Visualization, User Experience.**

## I. INTRODUCTION

Streamlining development processes is a challenging task for organizations in the current fast-paced commercial world. There is a greater need than ever for effective solutions to manage critical operations due to the increasing complexity of operations and the requirement for real-time insights. Unfortunately, current solutions frequently fall short because they don't have the user-centric design, scalability, or integration that modern organizations require to fulfill their changing needs for Business Dashboard [7].

A major obstacle that corporate development teams encounter is the absence of an all-inclusive administrative dashboard capable of efficiently centralizing and streamlining processes. Though they frequently have fragmented interfaces, limited functionality, and compatibility problems with other systems, existing dashboards may provide some partial answers. This fragmented strategy not only reduces output but also makes it more difficult to make decisions by restricting access to real-time data and insights.

This paper suggests creating and implementing a comprehensive admin dashboard that is powered by the Refine framework, Ant Design, and GraphQL API to handle these urgent concerns. Our solution intends to transform enterprise development processes by providing a seamless combination of an intuitive user interface, real-time data visualization, and sophisticated user management operations, all while utilizing the powerful potential of these technologies.

Moreover, the benefits of the Refine framework[9] are evident when contrasting our suggested solution with alternative technologies like conventional React or Node.js frameworks, as well as well-known platforms like Salesforce CRM. Although React and Node.js are widely used and versatile, they frequently need significant customization and integration work to reach the degree of coherence and functionality that Refine offers right out of the box.

Similar to this, Salesforce CRM may not offer the flexibility and customization options required to satisfy the particular needs of enterprise development processes, despite its robust customer relationship management functions. Refine, on the other hand, provides a flexible GraphQL API and a modular architecture that enables smooth adaptability to changing business demands without being constrained by a pre-packaged solution.

Furthermore, Refine's benefits are very evident when compared to other programming languages and frameworks. Because of its extensive feature set and user-friendly design, enterprise teams can concentrate on innovation instead of tedious technical problems, saving a substantial amount of time and effort during development. Furthermore, Refine offers a degree of security and compliance that can be absent from other languages or frameworks because of its strong user management and access control features.

Our suggested admin dashboard, which is enabled by GraphQL API, Ant Design, and Refine, essentially signifies a revolution in corporate development procedures. Our goal is to provide businesses with a solution that not only addresses their immediate needs but also establishes the groundwork for future growth and innovation by utilizing the distinct advantages of these technologies. The opportunities are boundless and the route to business success is more obvious than ever with Refine.

## II. LITERATURE SURVEY

A thorough approach is necessary when implementing an enterprise monitoring dashboard. Dashboards provide a consolidated view of key performance indicators (KPIs), which are essential for directing company decisions. They function as executive intranets. Nevertheless, careful planning, designing, and deploying are necessary for a successful implementation. This procedure includes gathering and synthesizing pertinent information, making sure strategy is in line, and taking technology into account. Project success requires a disciplined technique, regardless of the technology selected. Our study attempts to give a thorough methodology for using dashboards successfully by synthesizing ideas from the literature and equipping businesses with useful information for well-informed decision-making [5].

To enhance our dashboard application we have used the Refine technology [9] which helps in developing a customized Dashboard helps in enhances the user-friendly, flexible dashboard for the customer or the end-user, as the technology has started emerging in recent years Refine is booming in the marketplace, it is embedded with the multiple technologies i.e.., the Refine Framework is a multilingual platform such as GraphQl [2], Ant Design UI [8], React Framework [3] and many more.

For our project, Ant Design UI [8] is essential since it helps create a user-friendly and effective admin panel. Its vast component library simplifies the construction of user interfaces while guaranteeing scalability and conformity to contemporary design standards.

Ant Design UI [10] proves to be a crucial tool with its extensive documentation and community assistance, helping us achieve our objective of creating an admin dashboard that is data-driven and tailored for enterprise use.

The Refine also provides smooth onboarding is ensured by the authentication procedure, which also offers strong password recovery and safe login and signup options. Granular access control regulates user behavior, protecting data and efficiently handling permissions. The home page includes a deals chart for business insights, real-time activity updates, and dynamic charts for important metrics. Complete CRUD functions, including field-specific search options and full profiles, are available for company management through the companies page. Real-time task updates and customization with multiple assignees and due dates are possible with the Kanban board. Personalized profile management options are available through account settings. A uniform user experience is maintained by the design, which guarantees complete responsiveness across devices [11].

# III. PROPOSED METHODOLOGY

Following a comprehensive analysis of current CRM systems to pinpoint common shortcomings and restrictions, such as a lack of adaptability, a poorly designed user interface, a lack of capabilities for data visualization, and laborious development procedures for creating a customer-centric CRM.

On the other hand, the technologies employed in this project have been emerging and combining in recent years, competing with other front-end languages.
The language such as React JS from [2] helped us to understand the difficulties in developing an application only using React JS, on research we found a blog on refine [9] where refine consists of various leveraging advantages over React and this framework helps us to achieve a dashboard in a customizable way.

This proposed methodology will demonstrate the implementation of key features such as User management and access control, Data visualization dashboards, CRUD operations on enterprise data, Real-time notifications and alerts, and Integration with existing enterprise systems.

The final deliverable will be a fully functional and deployable admin dashboard, along with documentation and guidelines for future enhancements. This project aims to contribute to the empowerment of enterprise development teams by providing a modern, efficient, and data-driven tool for managing crucial operations and gaining valuable insights

The proposed method has been implanted using the featured tech stack listed below:

      A. Proposed Architecture
      B. Refine Framework
      C. React Components
      D. Ant Design UI Library
      E. Graph QL API
      F. TypeScript Code

### A. *Proposed Architecture*

Following is the project architecture:

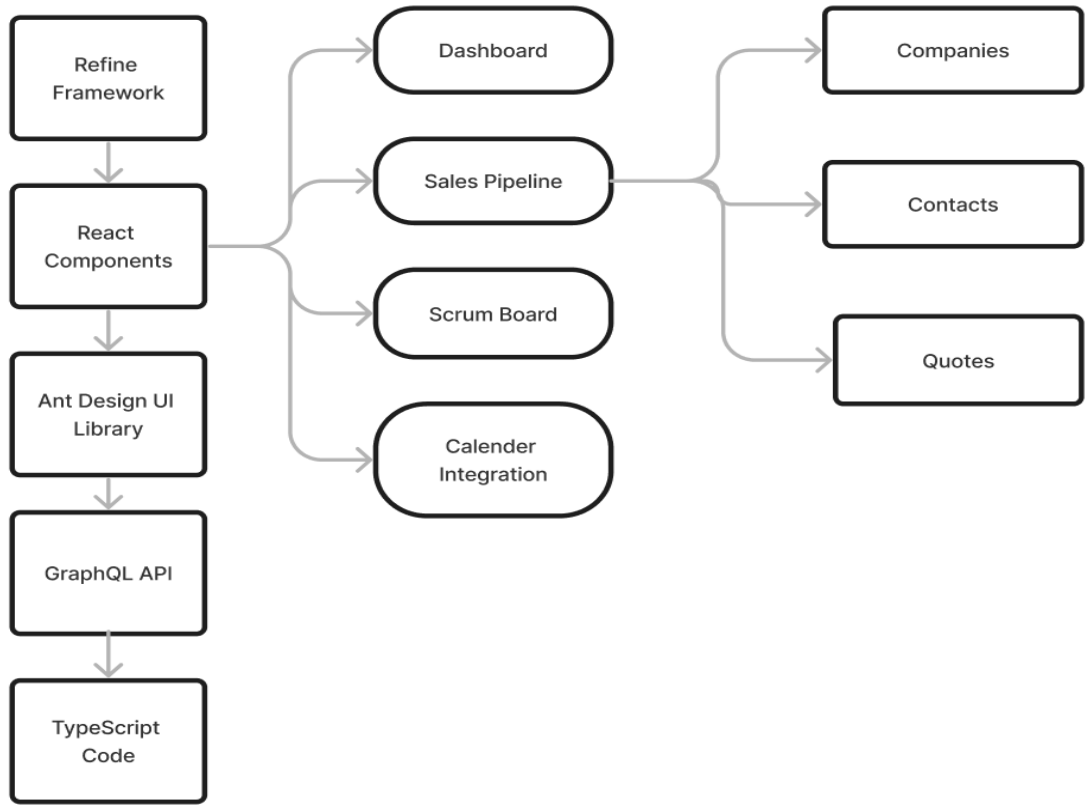

Fig. 1 Proposed Architecture

B. *Refine Framework*

An open-source React-meta framework, Refine is specifically designed to meet the needs of enterprise development teams. As part of the core package, you get data handling, authentication, access control, and other app requirements simplified and various functionalities as it is a traditional framework developed in recent years. You can create your internal tools, admin panels, dashboards, B2B applications, and any kind of CRUD application since you are production-ready. putting it into practice straight away or using it as a coding base.

As mentioned the Refine framework is the most booming framework with various technical features embedded in it. As a result of low/no-code solutions, this shortcoming is addressed, but a new set of challenges arises, it overcomes the listed challenges,

- Locking in vendors
- Style & customization options are limited
- Inexperienced developers
- Support for complex use cases is limited

C. *React Components*

The enterprise admin dashboard is built using React components, which provide modularity and code organization. Components streamline development cycles by reducing reusability and composability, leading to consistency across applications. To ensure dashboard functionality, we leverage Reacts real-time updates and lifecycle management capabilities. Performance enhancements ensure responsiveness by integrating

UI libraries like Ant Design. React components thus play a pivotal role in crafting a scalable, efficient, and visually appealing admin dashboard tailored to the needs of modern enterprises.

Real-time data (raw data) is collected in large quantities from the business organization. The raw data includes information of rides and zones. The data is then imported into a data frame and the columns and information present within the dataset are analysed. This feature helps us to divide the page into various components such as upcoming events block, graphical analysis profit block, and much more are this technology [2] has helped us to analyze the libraries in React JS to build various components in a dashboard.

The React components consist of the following:

- Dashboard
- Sales Pipeline
- Scrum Board

Clear Sales Pipelines supports the Monitor opportunities, visualizing stages, and closing deals more efficiently.

### D. Graph QL API

Our project's goal is to achieve efficient enterprise development through GraphQL's schema-based approach. Data insights gained from GraphQL can be used to empower enterprises to make more informed decisions, as demonstrated in performance evaluations. Our project's deployment is essential because GraphQL tooling is scalable and adaptable. To enable enterprise development, GraphQL APIs must be secured to ensure data integrity and user privacy. Using the insights we gathered, we aimed to leverage the strengths of GraphQL to empower enterprises through our admin dashboard solution.

API queries are executed using GraphQL using a type system you define for your data, along with a server-side runtime. Data and code are backed by your existing code and data, not a specific database or storage engine.

### D. Ant design UI library

Using Ant Design UI, you can build interactive user interfaces with easy-to-use React components. As well as being very easy to use, it is also very easy to integrate. Ant Design is one of the smart options and recently developed UI to design web applications using React. Using it is easy because it provides high-quality components.

### F. Type Script Code

In this venture, TypeScript code plays a vital part in upgrading the unwavering quality and practicality of the admin dashboard. By presenting inactive writing and sorting comments, TypeScript makes a difference capture mistakes amid improvement, diminishing the probability of runtime issues. Its solid writing framework guarantees consistency over the codebase, making it less demanding for engineers to collaborate and keep up the venture over time. Generally, TypeScript contributes to a stronger and adaptable arrangement, empowering smoother improvement and sending of the admin dashboard.

## IV. RESULT ANALYSIS

The result analysis will show the expected results of our designed and developed dashboard by displaying the various figures.

1. *DASHBOARD ACCOUNT PAGE:*

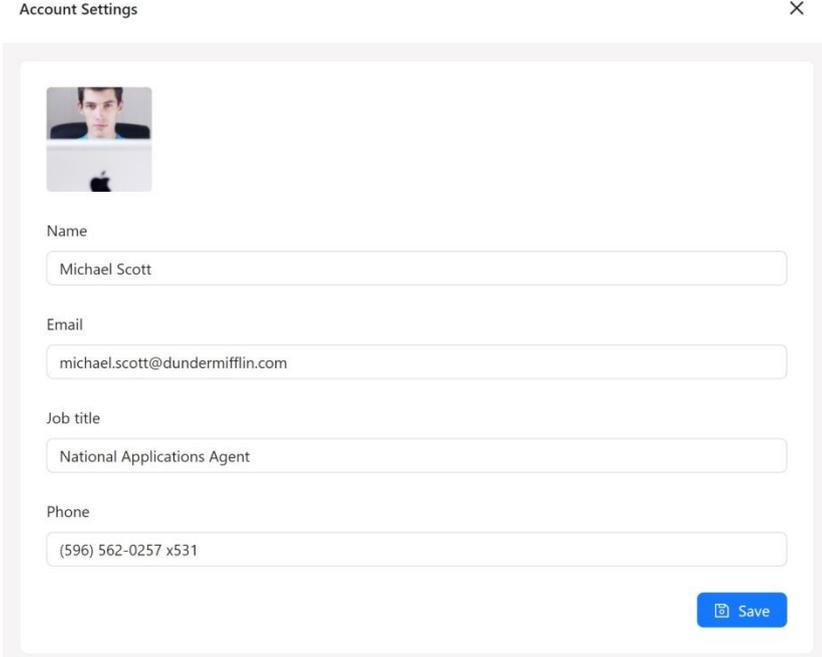

Fig 2. Account settings for the CRM page

The above figures tell us about the account settings of the logged-in user and the specific details of the user who is using the application (Dashboard).

2. *Dashboard Page:*

   This is the first page which contains the analytics of the client company, The dashboard is the UI which consists of the information about the enterprise activities and planning process following details:
   1) Number of clients dealing with an enterprise.
   2) Number of contacts in the database
   3) Total deals in the sales pipeline.
   4) Upcoming events of the enterprise.
   5) Deals chart which displays the revenue and expenditure of the company.

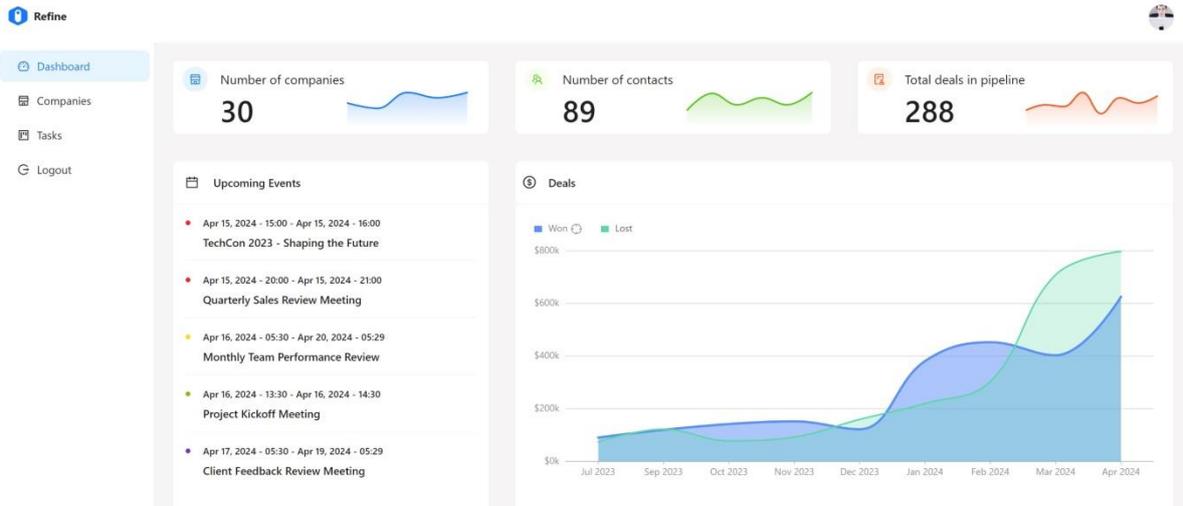

Fig 3. Dashboard.

Clear Sales Pipelines: Monitor opportunities, visualize stages, and close deals more efficiently.

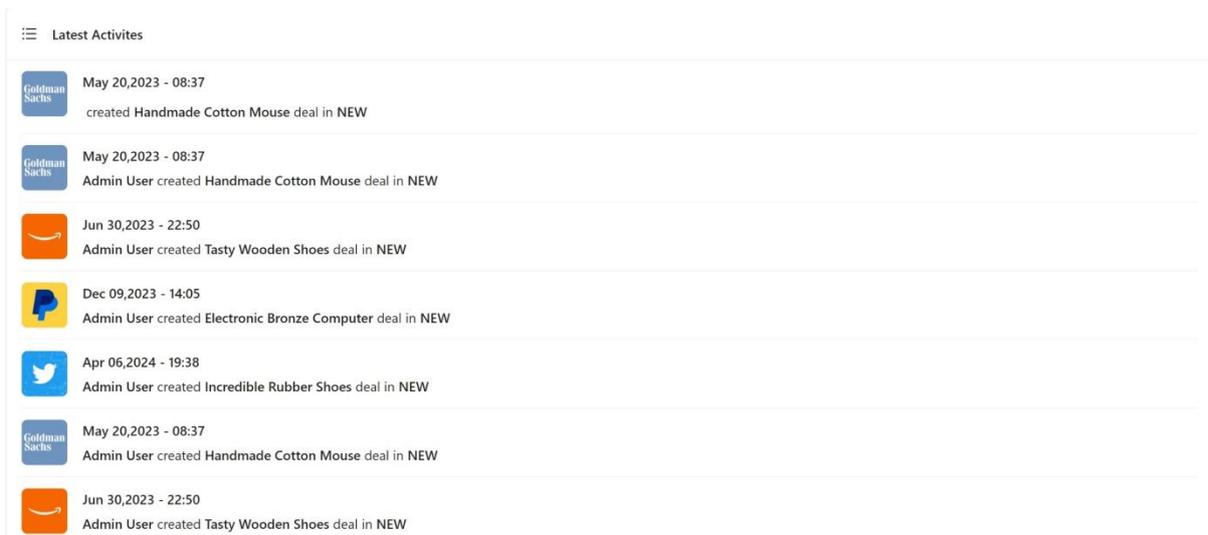

Fig 3.1. Dashboard- Latest Activities

In the above page of the application, it displays the recent or post activities in chronological order.

3. **Company Page:**

   The second page, the company's page, has information about the businesses that the firm has dealt with the enterprise. It displays the client's details that has made an agreement or deal with the enterprise for business.
    Included is the following data:

   1) Title of Company
   2) Amount of Open Deals: The deal value with the Company
   3) Actions: Edit and Delete are the two accessible actions.
   
   The figure represented below displays the above-mentioned data.

Fig 4. Company's or Clients Details Page.

4. *Creation of Client Lead details:*

Fig 5. A page for Creating Company Lead.

We also have the option to create a new company lead or a client lead which consists the following:
1) Details: Company name ,
2) Sales owner.

This is used to create new deals with the company, whenever the a new deal is done with the enterprise that is displayed over here. The above figure displays the details of the company deal.

**4.1.** *Modifying the created company or client details:*

This option or feature helps to edit the new company details. As u can see we can edit the sales owner, industry, revenue, and country which it belongs to, And We can see who is handling the deal between the client and enterprise is the Lead Handler, The Lead Handler helps to edit or modify the details of the client or he can also be called as the Administrator.

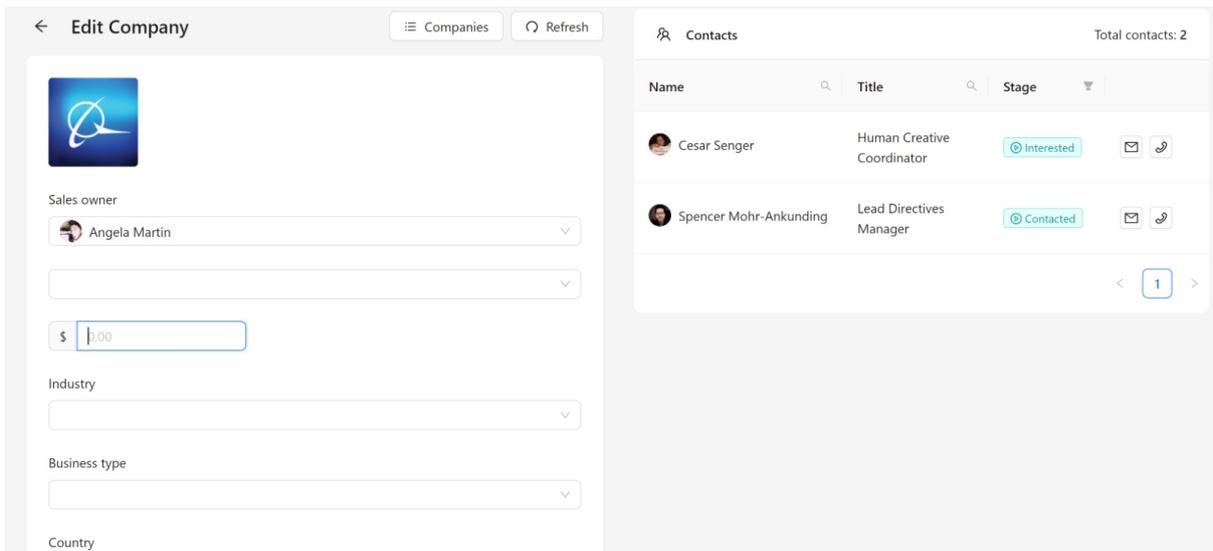

Fig 5.1. Modifying the client Page.

5. *Kanban Boards:*

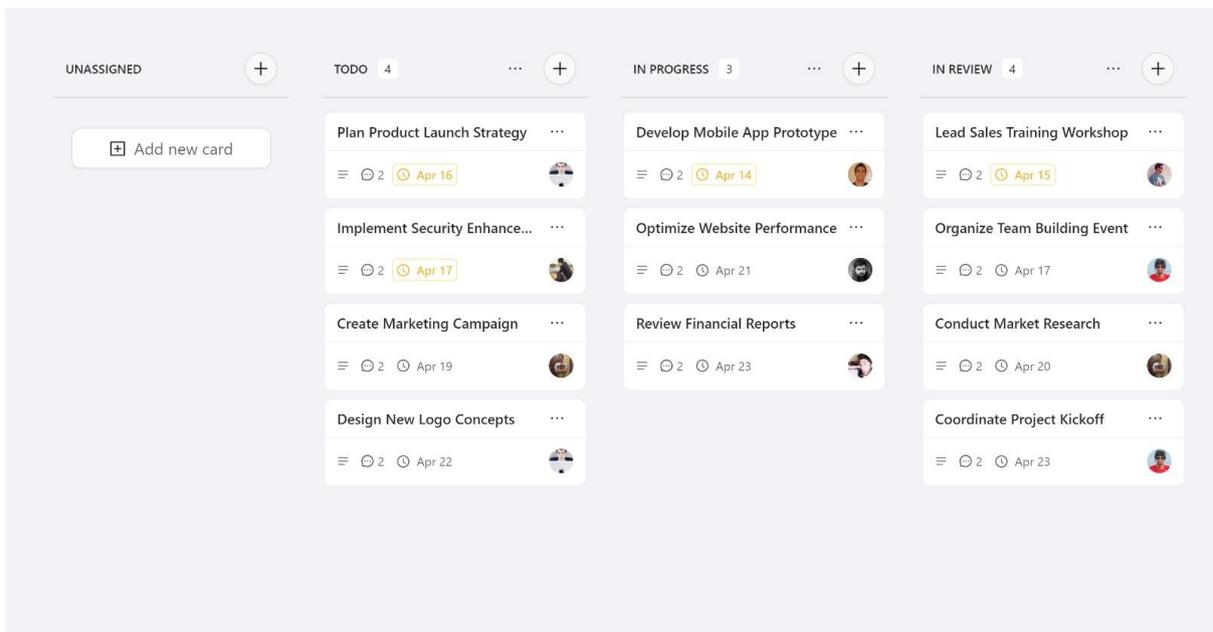

Fig 6. Kanban Boards.

The Kanban board lets us manage the upcoming tasks according to the deadline with a drag-and-drop feature and has real-time customization, It is similar to any ticketing too with a customization feature for any enterprise.

## V. CONCLUSION

In conclusion, an important step forward in business development processes has been made with the creation and implementation of the extensive admin dashboard using Refine, React components, Ant Design, GraphQL API, and TypeScript code. By carefully analyzing the shortcomings of the existing CRM system and conducting a comparison analysis, this project has proven the superiority of the suggested technique. The admin dashboard provides a revolutionary solution for businesses by tackling important issues including flexibility, user interface design, data visualization, and development efficiency. Case studies and prototyping validation have highlighted the approach's efficacy and usefulness. In the future, the project creates opportunities for more study and advancement, indicating that corporate management systems will continue to innovate. In the end, the admin dashboard gives corporate development teams access to a cutting-edge, effective, and data-driven solution for handling operations and driving success.